\documentclass[prd,showpacs,preprintnumbers,amsmath,amssymb]{revtex4}

\usepackage{graphicx}
\usepackage{dcolumn}
\usepackage{bm}

\def\be{\begin{equation}}
\def\ee{\end{equation}}
\def\bea{\begin{eqnarray}}
\def\eea{\end{eqnarray}}
\def\ba{\begin{array}}
\def\ea{\end{array}}

\begin{document}

\preprint{UTHET-10-0201}

\title{Analytic calculation of properties of holographic superconductors}

\author{George Siopsis}%
 \email{siopsis@tennessee.edu}
\author{Jason Therrien}%
 \email{jtherrie@utk.edu}
\affiliation{%
Department of Physics and Astronomy,
The University of Tennessee,
Knoxville, TN 37996 - 1200, USA.
}%

\date{February 2010}

\begin{abstract}
We calculate analytically properties of holographic superconductors in the probe limit.
We analyze the range $1/2 < \Delta < 3$, where $\Delta$ is the dimension of the operator that condenses.
We obtain the critical temperature in terms of a solution to a certain eigenvalue problem.
Near the critical temperature, we apply perturbation theory to determine the temperature dependence of the condensate.
In the low temperature limit we show that the condensate diverges as $T^{-\Delta/3}$ for $\Delta < 3/2$ whereas it asymptotes to a constant value for which we provide analytic estimates for $\Delta > 3/2$.
We also obtain the frequency dependence of the conductivity by solving analytically the wave equation of electromagnetic perturbations.
We show that the real part of the DC conductivity behaves as $e^{-\Delta_g /T}$ and estimate the gap $\Delta_g$ analytically. Our results are in good agreement with numerical results.

\end{abstract}

\maketitle

\section{Introduction}

The BCS theory of superconductivity \cite{PhysRev.108.1175} is a microscopic theory of weakly coupled superconductors and describes them with great accuracy.
However in the last few decades it has been discovered that there is another class of superconductors that are strongly coupled.
Conventional theories have failed to describe these systems.
The pairing mechanism and the normal state of the system before the onset of superconductivity remain open questions (see \cite{RevModPhys.66.763} for an introduction to strongly coupled superconductors).
In recent years it has been shown using the AdS$_{d+1}$/CFT$_d$ correspondence \cite{Maldacena:1997re} that for $d=3$ and $d=4$ we can create a strongly coupled superconducting system from its gravitational dual \cite{Gubser:2008px,Hartnoll:2008vx,Horowitz:2008bn,Franco:2009yz}, also including an external magnetic field \cite{Albash:2008eh,Hartnoll:2008kx}.
The $d=3$ systems are especially interesting since many strongly coupled superconductors are effectively $2+1-$dimensional  with the conductivity occurring in planes.
Phenomena such as the Hall effect
\cite{Hartnoll:2007ai} and Nernst effect \cite{Hartnoll:2007ih,
Hartnoll:2007ip, Hartnoll:2008hs} have also been shown to have dual gravitational
descriptions.
Other areas of Condensed Matter Physics have been discussed in terms of the AdS/CFT correspondence; see \cite{Hartnoll:2009sz} for a review.
The analysis for the most part has been done numerically with the exception of ref.~\cite{Konoplya:2009hv} in which the WKB approximation was employed in the calculation of conductivity.

Until recently the ground state of a holographic superconductor has been unreachable because a numerical solution to the non-linear field equations becomes increasingly cumbersome as the temperature approaches zero.
Progress was made in \cite{Horowitz:2009ij,Gubser:2009cg} where the ground state was reached numerically and the $T=0$ conductivity was analyzed.
Using the ground state, the Fermi surface of these systems was then explored \cite{Faulkner:2009am}.

Here we explore the properties of holographic superconductors using analytical techniques aiming at elucidating, among other things, the nature of the ground state.
We concentrate on the simplest case by working in the probe limit in which the
back reaction to the bulk metric can be ignored.
We obtain analytic approximations to the solutions of the non-linear field equations both near the critical temperature $T_c$ and in the low temperature limit
$T\to 0$.
This enables us to avoid problems with numerical instabilities which one encounters when one solves the field equations numerically near zero temperature.

The paper is organized as follows.
In section \ref{sec2} we review the field equations.
In section \ref{sec3} we discuss the properties of superconductors near the critical temperature.
In section \ref{sec4} we analyze the zero temperature limit.
In section \ref{sec5} we discuss the conductivity.
Finally, section \ref{sec6} contains
our concluding remarks.

\section{Field equations}
\label{sec2}

We are interested in the dynamics of a scalar field of mass $m$ coupled to a $U(1)$ vector potential in the backgound of a $3+1 -$~dimensional AdS Schwarzschild black hole with planar horizon of metric
\be ds^2 = - f(r) dt^2 + \frac{dr^2}{f(r)} + r^2 d{\vec x}^2 \ \ , \ \ \ \
f(r) = r^2 - \frac{r_+^3}{r} \ee
in units in which the AdS radius is $l=1$.
The radius of the horizon is $r_+$ and the Hawking temperature is
\be T = \frac{3}{4\pi}\ r_+ \ee
Assuming that the scalar field is of the form $\Psi (r)$ and the potential is an electrostatic scalar potential $\Phi (r)$, the
field equations are \cite{Hartnoll:2008vx}
\bea\label{eq1}
\Psi'' + \left( \frac{f'}{f} + \frac{2}{r} \right) \Psi' + \left( \frac{\Phi^2}{f^2} - \frac{m^2}{f} \right) \Psi &=& 0 \nonumber\\
\Phi'' + \frac{2}{r} \Phi' - 2\frac{\Psi^2}{f} \Phi &=& 0 \eea
Under the change of coordinates
\be z = \frac{r_+}{r} \ee
the field equations become
\bea
z\Psi'' - \frac{2+z^3}{1-z^3} \Psi' + \left[ z \frac{\Phi^2}{r_+^2 (1-z^3)^2} - \frac{m^2}{z(1-z^3)} \right] \Psi &=& 0 \nonumber\\
\Phi'' - \frac{2\Psi^2}{z^2 (1-z^3)} \Phi &=& 0
\eea
where prime now denotes differentiation with respect to $z$,
to be solved in the interval $(0,1)$, where $z=1$ is the horizon and $z=0$ is the boundary.

Near the boundary ($z\to 0$), we have approximately
\be\label{eq3} \Psi \approx \Psi^{(\pm)} z^{\Delta_\pm} \ \ , \ \ \ \ \Phi \approx \mu - \frac{\rho}{r_+} z \ee
where
\be \Delta_\pm = \frac{3}{2} \pm \sqrt{\frac{9}{4} + m^2} \ee
While a linear combination of asymptotics is allowed by the field equations, it turns out that any such combination is unstable \cite{Hertog:2004bb}.
However, if the horizon has negative curvature, such linear combinations lead to stable configurations in certain cases \cite{papa1}.

Thus, the system is labeled uniquely by the dimension $\Delta = \Delta_\pm$.
We shall examine the range
\be \frac{1}{2} < \Delta < 3 \ee
where $\Delta > 3/2$ ($\Delta < 3/2$) if $\Delta = \Delta_+$ ($\Delta = \Delta_-$),
corresponding to masses in the range $0>m^2> -9/4$ (above the Breitenlohner-Friedman bound \cite{Breitenlohner:1982jf,M-T}).

Demanding at the horizon
\be\label{eq7} \Phi (1) = 0 \ee
$\mu$ is interpreted as the chemical potential of the dual theory on the boundary.
$\rho$ is the charge density on the boundary and the leading coefficient in the expansion of the scalar yields vacuum expectation values of operators of dimension $\Delta_\pm$,
\be\label{eq4} \langle \mathcal{O}_{\Delta_\pm} \rangle = \sqrt 2 r_+^{\Delta_\pm} \Psi^{(\pm)} \ee
The field equations admit non-vanishing solutions for the scalar below a critical temperature $T_c$ where these operators condense.

\section{Near the critical temperature}
\label{sec3}

At the critical temperature $T_c$, $\Psi = 0$, so the field equation (\ref{eq1}) for the electrostatic potential reduces to $\Phi''=0$. We may set
\be \Phi (z) = \lambda r_{+c} (1-z) \ \ , \ \ \ \ \lambda = \frac{\rho}{r_{+c}^2} \ee
where $r_{+c}$ is the radius of the horizon at $T=T_c$.

As $T\to T_c$, the field equation for the scalar field $\Psi$ approaches the limit
\be - \Psi'' + \frac{2+z^3}{z(1-z^3)} \Psi' + \frac{m^2 }{z^2(1-z^3)} \Psi = \frac{\lambda^2 }{(1+z+z^2)^2} \Psi \ee
which is valid even at $T=T_c$, because it is linear in $\Psi$
(one may divide $\Psi$ by the leading coefficient in the expansion around $z=0$ before letting $\Psi\to 0$).

To match the behavior at the boundary, define
\be\label{eq2} \Psi (z) = \frac{\langle \mathcal{O}_\Delta \rangle}{\sqrt 2 r_+^\Delta} z^\Delta F(z) \ee
where we used eq.~(\ref{eq4}) to express the leading order coefficient so that $F$ is normalized to $F(0)=1$.

We deduce
\be\label{eq5b} - F'' + \frac{1}{z} \left[ \frac{2+z^3}{1-z^3} -2\Delta \right] F' + \frac{\Delta^2 z }{1-z^3} F = \frac{\lambda^2 }{(1+z+z^2)^2} F \ee
to be solved subject to the boundary condition
\be F' (0) = 0 \ee
The eigenvalue $\lambda$ minimizes the expression
\be\label{eq5a} \lambda^2 = \frac{\int_0^1 dz\ z^{2\Delta -2} \{ (1-z^3) [F'(z)]^2 + \Delta^2 z [F(z)]^2 \} }{\int_0^1 dz \ z^{2\Delta -2} \frac{1-z}{1+z+z^2} [F(z)]^2} \ee
To estimate it, use the trial function
\be\label{eq5} F= F_\alpha (z) \equiv 1 - \alpha z^2 \ee
For $\Delta =1$ we obtain
\be \lambda_\alpha^2 = \frac{6-6\alpha + 10\alpha^2}{2\sqrt 3 \pi - 6\ln 3
+ 4 (\sqrt 3 \pi +3\ln 3 - 9)\alpha + (12\ln 3 - 13)\alpha^2} \ee
which attains its minimum at $\alpha \approx 0.24$. We obtain
\be \lambda^2 \approx \lambda_{0.24}^2 \approx 1.27 \ee
to be compared with the exact value $\lambda^2 = 1.245$.
The critical temperature is
\be\label{eqTc} T_c = \frac{3}{4\pi} r_{+c} = \frac{3}{4\pi} \sqrt{\frac{\rho}{\lambda}} \ee
so for $\Delta = 1$, $T_c \approx 0.225\sqrt\rho$, in very good agreement with the exact $T_c = 0.226\sqrt\rho$ \cite{Hartnoll:2008vx}.

\begin{figure}
\includegraphics{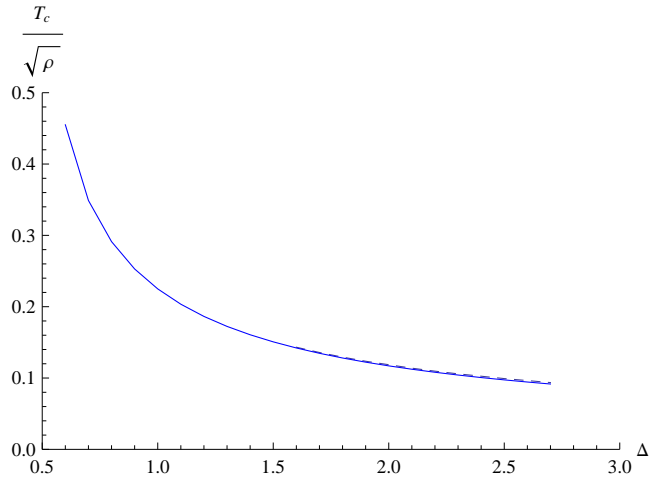}
\caption{The critical temperature in units of $\sqrt\rho$ {\em vs.} the dimension of the condensate~$\Delta$ (eq.~(\ref{eqTc})).
The solid line is the analytic estimate (eqs.~(\ref{eq5a}) and (\ref{eq5})) and the almost indistinguishable dashed line is found by solving the field eqs.~(\ref{eq1}) numericallly.}
\label{fig3}
\end{figure}

Similarly, for $\Delta=2$ we obtain
\be \lambda_\alpha^2 = 2 \frac{1 - \frac{4}{3}\alpha + \frac{4}{5} \alpha^2}{3-\ln 3 - \frac{\pi}{\sqrt 3} + (\frac{13}{3} - 4\ln 3)\alpha + ( \frac{\pi}{\sqrt 3} - \frac{7}{10} + \ln 3)\alpha^2} \ee
whose minimum is
$\lambda^2 \approx 17.3$ (at $\alpha\approx 0.6$) to be compared with the exact value $\lambda^2 = 16.754$.
The critical temperature in this case is $T_c \approx 0.117\sqrt\rho$, in very good agreement with the exact $T_c = 0.118\sqrt\rho$ \cite{Hartnoll:2008vx}.
In fig.~\ref{fig3} we compare the analytic estimate of the critical temperature obtained from eqs.~(\ref{eq5a}) and (\ref{eq5}) with exact numerical results.
The agreement between the two is excellent.

Away from (but close to) the critical temperature, the field equation (\ref{eq1}) for $\Phi$ becomes
\be
\Phi'' = \frac{\langle\mathcal{O}_\Delta\rangle^2}{2r_+^{2\Delta}}\ \frac{z^{2(\Delta -1)} F^2(z)}{1-z^3}\ \Phi \ee
where the parameter $\langle\mathcal{O}_\Delta\rangle^2/(2r_+^{2\Delta})$ is small.
We may expand in the small parameter,
\be \frac{\Phi}{r_+} = \lambda (1-z) + \frac{\langle\mathcal{O}_\Delta\rangle^2}{2r_+^{2\Delta}} \chi (z) + \dots \ee
We deduce for the correction $\chi$ near the critical temperature
\be
\chi'' = \lambda \frac{z^{2(\Delta -1)} F^2(z)}{1+z+z^2} \ee
with $\chi(1) = \chi'(1)=0$.

\begin{figure}
\includegraphics{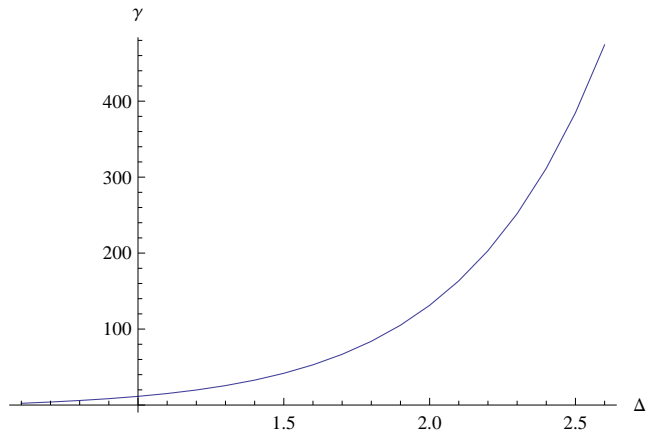}
\caption{The parameter $\gamma$ that determines the condensate near the critical temperature {\em vs.} the dimension of the condensate~$\Delta$ (eq.~(\ref{eqgam})).}
\label{fig1}
\end{figure}

To find the temperature, we need
\be\label{eq6} \chi_1' (0) = \lambda \mathcal{C} \ \ , \ \ \ \
\mathcal{C} = \int_0^1 dz \ \frac{z^{2(\Delta -1)} F^2(z)}{1+z+z^2} \ee
From eq.~(\ref{eq3}), we deduce the ratio
\be \frac{\rho}{r_+^2} = \lambda \left( 1 +  \frac{\mathcal{C} \langle\mathcal{O}_\Delta\rangle^2}{2r_+^{2\Delta}} + \dots \right) \ee
therefore the condensate near the critical temperature is
\be\label{eqgam} \langle \mathcal{O}_\Delta \rangle \approx
\gamma
T_c^\Delta \left( 1 - \frac{T}{T_c} \right) \ \ , \ \ \ \ \gamma =
\frac{2}{\sqrt{\mathcal{C}}} \left( \frac{4\pi }{3} \right)^{\Delta}
\ee
Using the trial functions (\ref{eq5}), for $\Delta =1$, we obtain from (\ref{eq6}), $\mathcal{C} \approx 0.54$ and $\gamma \approx 11.4$ to be compared with the exact $\gamma = 9.3$ \cite{Hartnoll:2008vx}.
Similarly, for $\Delta =2$, we find $\mathcal{C} \approx 0.07$ and $\gamma \approx 133 $ to be compared with the exact $\gamma = 144 $ \cite{Hartnoll:2008vx}.
In fig.~\ref{fig1} we plot the analytic prediction (\ref{eqgam}) for the parameter $\gamma$ as a function of the dimension of the condensate $\Delta$.
Notice that $\gamma$ diverges as $\Delta\to 3$.

\section{Low temperatures}
\label{sec4}

Turning to low temperatures,
as $T\to 0$, we expect a simple scaling, $\Psi = \Psi (bz)$, $\Phi = \Phi (bz)$ where $b\to\infty$.
Then scaling $z\to z/b$ and letting $b\to\infty$, the field equations (\ref{eq1}) simplify, since the dominant contribution comes from the neighborhood of the boundary ($z=0$).
Thus at low temperature
we obtain the simplified system of equations
\bea\label{eq28} -F'' + \frac{2(1-\Delta)}{z} F' - \frac{\Phi^2}{r_+^2 } F &=& 0 \nonumber\\
\Phi'' - \frac{\langle\mathcal{O}_\Delta\rangle^2}{r_+^{2\Delta}} F^2z^{2(\Delta-1)} \Phi &=& 0 \eea
where we restored the original coordinate $z$ (before scaling).

This system of coupled non-linear equations is to be solved subject to the boundary condition at the horizon
\be\label{eq29} 3F'(1) + \Delta^2 F(1) = 0 \ee
which is obtained from (\ref{eq1}), (\ref{eq7}) and (\ref{eq2}),
as well as those at the boundary, $F(0)=1$ and $F'(0)=0$.

For $\Delta = \Delta_- < 3/2$, numerical results indicate that
$F\to 1$ as $T\to 0$. The solution to the field equation (\ref{eq28}) for $\Phi$ is
\be\label{eq30} \Phi (z) = \mathcal{A} r_+ \sqrt z K_{\frac{1}{2\Delta}} ( b^\Delta z^\Delta ) \ \ , \ \ \ \ b^\Delta = \frac{\langle\mathcal{O}_\Delta\rangle }{\Delta r_+^\Delta} \ee
The other solution is rejected because it is large at the horizon, contradicting the boundary condition (\ref{eq7}).
Notice that at the horizon $\Phi (1) \sim e^{-b^\Delta}$, which is an exponentially small error in the $T\to 0$ ($b\to\infty$) limit.

From the behavior of $\Phi$ (eq.~(\ref{eq30})) near the boundary ($z=0$), using (\ref{eq3}), we deduce
\be\label{eq31} \frac{\rho}{r_+^2} = - \frac{\Gamma (-\frac{1}{2\Delta})}{2^{1+\frac{1}{2\Delta}}} \mathcal{A} b \ee
The field equation (\ref{eq28}) for $F$ becomes
\be -F'' + \frac{2(1-\Delta)}{z} F' - \mathcal{A}^2 bz (K_{\frac{1}{2\Delta}} ( b^\Delta z^\Delta ) )^2 F = 0 \ee
Using perturbation theory, we obtain the solution
\be F(z) = 1 - \mathcal{A}^2 b \int_0^z dz'\ {z'}^{2(1-\Delta)} \int_0^{z'} dz''\ z'' (K_{\frac{1}{2\Delta}} ( b^\Delta {z''}^\Delta ) )^2 \ee
Applying the boundary condition (\ref{eq29}) at the horizon, we obtain
\be \mathcal{A}^2 = \frac{\Delta^2 b^{2(2-\Delta)}}{3b\mathcal{F}_\Delta' (b) + \Delta^2 \mathcal{F}_\Delta (b)} \ \ , \ \ \ \ \mathcal{F}_\Delta ( x) = \int_0^x dx'\ {x'}^{2(1-\Delta)} \int_0^{x'} dx''\ x'' (K_{\frac{1}{2\Delta}} ( {x''}^\Delta ) )^2 \ee
For large $b$, the denominator scales like $b^{3-2\Delta}$, showing that $\mathcal{A} \sim \sqrt b$. Then eq.~(\ref{eq31}) implies $\rho/r_+^2 \sim b^{3/2}$.
Since
$\langle\mathcal{O}_\Delta\rangle \sim (r_+b)^\Delta$ (see eq.~(\ref{eq30})), it follows that
\be \frac{\langle \mathcal{O}_\Delta \rangle^{1/\Delta}}{T_c} \sim b^{1/4} \sim \left( \frac{T}{T_c} \right)^{-1/3} \ee
showing that the condensate diverges as $\langle\mathcal{O}_\Delta \rangle \sim T^{-\Delta/3}$ for $\Delta$ in the range $1/2 < \Delta < 3/2$ signaling the breakdown of the probe-limit approximation at low temperatures.

Turning to the case $\Delta = \Delta_+ > 3/2$, notice that as
we switch from $\Delta=\Delta_-$ to $\Delta = \Delta_+$, the boundary conditions at $z=0$ change, but not at the horizon. Thus, for a given $m$, the electrostatic potential $\Phi$ has the same asymptotic behavior for both $\Delta_+$ and $\Delta_-$.
In terms of the scalar field, this implies $F\approx 1$ near the boundary ($z=0$), whereas
%
\be\label{eq36} F(z) \approx \left( \frac{b z}{\alpha} \right)^{3-2\Delta} \ee
asymptotically ($z\gtrsim 1/b$, where $b\gg 1$ is to be determined).
Then in the asymptotic regime, eq.~(\ref{eq28}) for $\Phi$ has solution ({\em cf.}~eq.~(\ref{eq30}))
\be\label{eq37} \Phi (z) = \mathcal{A} r_+ \sqrt{bz} K_{\frac{1}{2(3-\Delta)}} ( b^{3-\Delta} z^{3-\Delta} ) \ \ , \ \ \ \
b^\Delta = \frac{\langle\mathcal{O}_\Delta\rangle \alpha^{2\Delta-3}}{ (3-\Delta) r_+^\Delta} \ee
Notice that a singularity appears to develop at $\Delta = 3$ indicating the onset of a quantum phase transition.

Near the boundary, we deduce the estimate for the ratio $\rho/r_+^2$,
\be\label{eq38} \frac{\rho}{r_+^2} = - \frac{\Gamma ( -\frac{1}{2(3-\Delta)})}{2^{1+\frac{1}{2(3-\Delta)}}} \mathcal{A} b \ee
However, this is not a good estimate. In fact, it diverges at $\Delta = 5/2$.
We shall improve on this estimate by better accounting for the behavior of $F$ near the boundary (where eq.~(\ref{eq36}) ought to be replaced by $F\approx 1$).

The field equation (\ref{eq28}) for $F$ is
\be\label{eq39} -F'' + \frac{2(1-\Delta)}{z} F' - \mathcal{A}^2 bz (K_{\frac{1}{2(3-\Delta)}} ( b^{3-\Delta} z^{3-\Delta} ) )^2 F = 0 \ee
Unlike with $\Delta < 3/2$, in this case the term involving the electrostatic potential $\Phi$ cannot be treated as a perturbation. By rescaling $z \to z/b$, eq.~(\ref{eq39}) becomes
\be\label{eq40} -F'' + \frac{2(1-\Delta)}{z} F' - \hat{\mathcal{A}}^2 z (K_{\frac{1}{2(3-\Delta)}} ( z^{3-\Delta} ) )^2 F = 0 \ \ , \ \ \ \ \hat{\mathcal{A}} = \frac{\mathcal{A}}{b} \ee
With
\be \mathcal{A} \sim b \ee
eq.~(\ref{eq40}) is independent of temperature and provides a good approximation to $F$ and the corresponding eigenvalue $\hat{\mathcal{A}}$ at $T=0$.
It ought to be solved in the interval $(0,\infty)$ subject to the boundary conditions $F(0)=1$, $F'(0)=0$, $F\to 0$ as $z\to\infty$ (see eq.~(\ref{eq36})).
These conditions determine the eigenvalue $\hat{\mathcal{A}}$.

\begin{figure}
\includegraphics{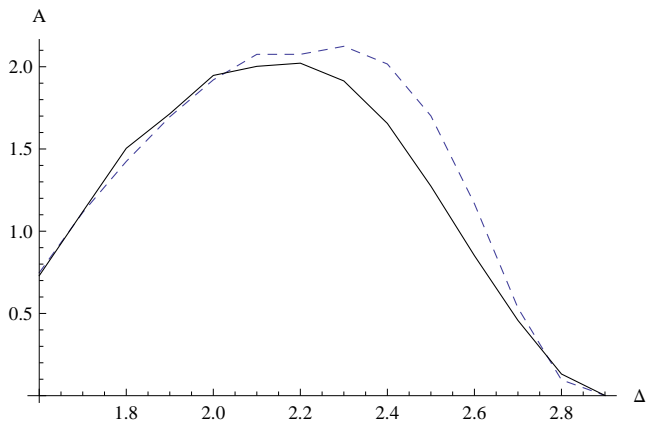}
\caption{The eigenvalue $\hat{\mathcal{A}}$ defined in (\ref{eq40}) as a function of $\Delta$ [solid line] compared with the estimate (\ref{eq40a}) [dashed line].}
\label{fig2}
\end{figure}

\begin{figure}
\includegraphics{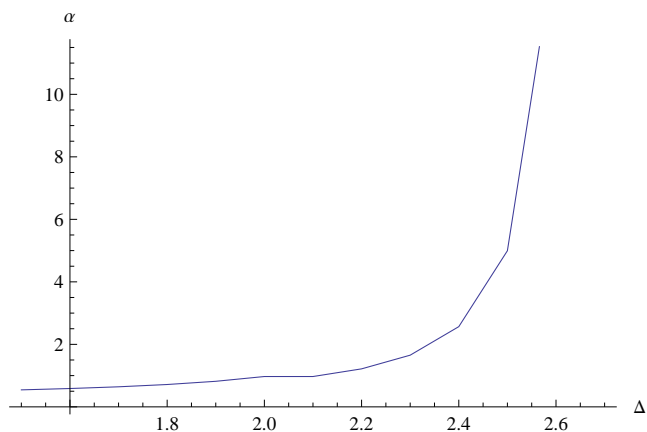}
\caption{The parameter $\alpha$ that determines the approximation (\ref{eq40b}) to the scalar field as a function of $\Delta$.}
\label{fig2a}
\end{figure}

To estimate $\hat{\mathcal{A}}$,
note that
\be\label{eq40a} \hat{\mathcal{A}}^2 = \frac{\int_0^\infty dz\ z^{2(\Delta -1)} [F'(z)]^2}{\int_0^\infty dz\ z^{2\Delta -1} [K_{\frac{1}{2(3-\Delta)}} (z^{3-\Delta}) F(z)]^2} \ee
The eigenfunction $F(z)$ minimizes this expression. We may substitute the trial function
\be\label{eq40b} F_\alpha(z) = \left( \frac{\alpha}{z} \right)^{2\Delta -3} \tanh \left( \frac{z}{\alpha} \right)^{2\Delta -3} \ee
which obeys the correct boundary conditions.
It interpolates smoothly between a constant value ($F=1$) near the boundary and the power behavior (\ref{eq36}) away from the boundary. The parameter $\alpha$ is fixed by minimizing the ratio (\ref{eq40a}). Similar functions have been considered before \cite{Gregory:2009fj,Pan:2009xa} but without a determination of the parameter $\alpha$.
In fig.~\ref{fig2} we compare the analytic estimate of the eigenvalue $\hat{\mathcal{A}}$ using the trial functions (\ref{eq40b}) with exact numerical results.
The agreement between the two is excellent. Notice that the limit $\Delta\to 3$ is singular. $\hat{\mathcal{A}} \to 0$ in this limit. Also, the parameter $\alpha$ labeling the trial function that minimizes (\ref{eq40a}), which is plotted in fig.~\ref{fig2a} as a function of the dimension $\Delta$, diverges as $\Delta\to 3$.


For $\Delta =2$ the minimum is obtained for $\alpha \approx 0.8$ which yields
$\hat{\mathcal{A}} \approx 1.92$.
In this case, $\hat{\mathcal{A}}$ can be found exactly because eq.~(\ref{eq40}) can be solved analytically.
We find explicitly
\be F(z) = \frac{\pi}{2z} \left[ Y_0 \left( \sqrt{\frac{\pi}{2}} \hat{\mathcal{A}} \right)J_0 \left( \sqrt{\frac{\pi}{2}} \hat{\mathcal{A}} e^{-z} \right) - J_0 \left( \sqrt{\frac{\pi}{2}} \hat{\mathcal{A}} \right)Y_0 \left( \sqrt{\frac{\pi}{2}} \hat{\mathcal{A}} e^{-z} \right) \right] \ee
which obeys the correct boundary conditions at $z=0$. Demanding $F\to 0$ as $z\to\infty$ then yields
\be J_0 \left( \sqrt{\frac{\pi}{2}} \hat{\mathcal{A}} \right) = 0 \ee
showing that
\be\label{eq46} \hat{\mathcal{A}} = \frac{\mathcal{A}}{b} = \xi_0 \sqrt{\frac{2}{\pi}} \ \ , \ \ \ \ \xi_0 \approx 2.4 \ee
where $\xi_0$ is the first root of the Bessel function $J_0$.
Numerically, $\hat{\mathcal{A}} = 1.91$, in good agreement with our earlier estimate.

Restoring $b$,
\be F(z) = \frac{\pi}{2bz} Y_0 (\xi_0) J_0 (\xi_0 e^{-bz} ) \ee
so $F(z) = 1 + \mathcal{O} (z^2)$, as desired. Away from the boundary
($z\gtrsim 1/b$),
\be F(z) \approx \frac{\pi Y_0 (\xi_0)}{2b z} \ee
which upon comparison with (\ref{eq36}) yields
\be \alpha = \frac{\pi Y_0 (\xi_0)}{2} \approx 0.8
\ee
in excellent agreement with our earlier estimate.

To calculate the condensate at $T=0$, we need a better estimate of the ratio (\ref{eq38}).
To this end, we shall solve eq.~(\ref{eq28}) for $\Phi$ perturbatively using (\ref{eq37}) with $\Delta =2$ as our zeroth-order solution. We obtain
\be \Phi (z) = \sqrt{\frac{\pi}{2}} \mathcal{A} r_+ \ e^{-bz} \left[ 1
+ \frac{\alpha}{2} - \frac{\pi\alpha^2}{\sin\pi\alpha} e^{2bz} + \frac{\alpha}{2} F(-\alpha, 1; 1-\alpha; -e^{2bz/\alpha})
+ \frac{\alpha^2}{2(1-\alpha)} e^{2bz/\alpha} F(1-\alpha, 1, 2-\alpha; -e^{2bz/\alpha} )
\right] \ee
We deduce the ratio
\be \frac{\sqrt\rho}{r_+} \approx b\sqrt{\xi_0} \left[ 1 -\frac{\delta}{2} \right] \ \ , \ \ \ \
\delta = \left. \alpha + \frac{\pi\alpha^2}{\sin\pi\alpha} + \frac{\alpha^2}{2} \psi \left( \frac{1-\alpha}{2} \right) - \frac{\alpha^2}{2} \psi \left( - \frac{\alpha}{2} \right) \right|_{\alpha = \pi Y_0(\xi_0)/2} \ee
improving the zeroth-order result $\sqrt\rho/r_+ \approx \sqrt{\xi_0}\ b$ (eqs.~(\ref{eq38}) with $\Delta =2$ and (\ref{eq46})).
Numerically, $\delta = 0.58$
and the condensate at $T=0$ is
\be \frac{\sqrt{\langle\mathcal{O}_2\rangle}}{T_c} \approx \frac{1+\frac{\delta}{2}}{0.118\sqrt{\alpha\xi_0}} \approx 7.9 \ee
in good agreement with the exact result $\langle \mathcal{O}_2 \rangle^{1/2} = 8.3 T_c$.

\begin{figure}
\includegraphics{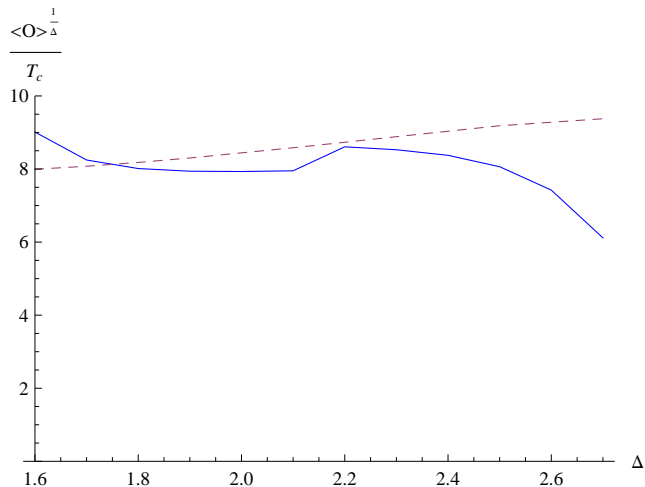}
\caption{The condensate at zero temperature as a function of $\Delta$. The solid line is our anaytic estimate whereas the dashed line is the exact numerical result.}
\label{fig5}
\end{figure}

To generalize to arbitrary $\Delta > 3/2$, substitute the approximation to the function $F_\alpha(z)$ (eq.~(\ref{eq40b})),
\be F_\alpha(z) \approx \left\{ \begin{array}{lll} 1 & , & z\le \alpha \\ (\alpha/z)^{2\Delta -3} & , & z>\alpha \end{array} \right. \ee
into the field equation for the electrostatic potential $\Phi$ and solve it to find a better approximation for $\Phi$ than (\ref{eq37}).
After rescaling, $z\to z/b$, for $z>\alpha$ we obtain
\be \Phi = \Phi_> (z) = \hat{\mathcal{A}} b r_+ \sqrt z K_{\frac{1}{2(3-\Delta)}} (z^{3-\Delta} ) \ee
whereas for $z\le \alpha$,
\be\label{eq38i} \Phi = \Phi_< (z) = \mathcal{B}_+ \sqrt z I_{\frac{1}{2\Delta}} \left( \frac{3-\Delta}{\Delta}\ \frac{z^\Delta}{\alpha^{2\Delta -3}} \right)
+ B_- \sqrt z I_{-\frac{1}{2\Delta}} \left( \frac{3-\Delta}{\Delta} \ \frac{z^\Delta}{\alpha^{2\Delta -3}} \right) \ee
providing the estimate for $\rho$ improving on (\ref{eq38}),
\be \rho = - \frac{\mathcal{B_+} br_+}{\Gamma ( 1 + \frac{1}{2\Delta})}
\left( \frac{3-\Delta}{2\Delta\alpha^{2\Delta-3}} \right)^{\frac{1}{2\Delta}} \ee
The coefficients $\mathcal{B}_\pm$ are found by matching the two expressions at $z=\alpha$.
For $\Delta =2$ we obtain
$\mathcal{B}_+ \approx -1.73 br_+$,
$\mathcal{B}_- \approx 1.96 br_+$,
therefore $\rho \approx 1.43 b^2 r_+^2$ and
$\sqrt{\langle \mathcal{O}_2 \rangle}/T_c \approx 7.9$, as before.
Fig.~\ref{fig5} shows the dependence of the condensate on the dimension $\Delta$. Our analytic estimate is in good agreement with the exact numerical value
for most of the range of $\Delta$. It becomes increasingly unreliable as $\Delta \to 3$.
This is expected from our estimate of the electrostatic potential (\ref{eq38i}) which is singular in the limit $\Delta\to 3$.

\section{Conductivity}
\label{sec5}

The conductivity on the boundary is found by applying a sinusoidal electromagnetic perturbation in the bulk of frequency $\omega$ obeying the
wave equation
\be\label{eq53} - \frac{d^2 A}{dr_*^2} + V A = \omega^2 A \ \ , \ \ \ \ V = 2f\Psi^2 \ee
where $A$ is any component of the perturbing electromagnetic potential along the boundary.
Eq.~(\ref{eq53}) is to be solved subject to ingoing boundary condition at the horizon
\be\label{eq54} A \sim e^{-i\omega r_*} \sim (1-z)^{-i\omega/(3r_+)} \ee
as $z\to 1$ ($r_*\to -\infty$), where $r_*$ is the tortoise coordinate
\be r_* = \int \frac{dr}{f(r)} = \frac{1}{6r_+} \left[ \ln \frac{(1-z)^3}{1-z^3}
-2\sqrt 3 \tan^{-1} \frac{\sqrt 3 z}{2+z} \right] \ee
with the integration constant chosen so that the boundary is at $r_* =0$.
We shall solve this equation in the entire frequency spectrum.

To this end, we shall replace the potential $V$ with its average $\langle V\rangle$ in a self-consistent manner.
We readily obtain the solution
\be A = e^{-i \sqrt{\omega^2 - \langle V\rangle}\ r_*} \ee
The other solution is discarded because it contradicts the boundary condition at the horizon (\ref{eq54}).
We deduce the conductivity
\be\label{eq57} \sigma(\omega) = \sqrt{ 1 - \frac{\langle V\rangle}{\omega^2}} \ee
The average value of the potential is found from
\be\label{eq58} \langle V \rangle = \frac{\int_{-\infty}^0 dr_* V |A(r_*)|^2}{\int_{-\infty}^0 dr_* |A (r_*)|^2} \ee
The integrals are well-defined if $\omega$ has an imaginary part (which should be set equal to zero at the end of the calculation).

For $\Delta < 3/2$, in the low temperature limit the potential simplies to
\be V \approx \frac{\langle \mathcal{O}_\Delta \rangle^2}{r_+^{2(\Delta-1)}} z^{2(\Delta -1)} (1-z^3) \ee
where we used eq.~(\ref{eq2}) with $F(z)\approx 1$.
Moreover, since $r_+\to 0$, the main contribution to the integrals in (\ref{eq58})
is from the vicinity of the boundary where $r_*\approx -z/r_+$. We deduce the leading contribution
\be\label{eq60} \langle V\rangle \approx \frac{\langle \mathcal{O}_\Delta \rangle^2}{r_+^{2(\Delta-1)}} \frac{\int_0^\infty dz \, z^{2(\Delta -1)} |A (-z/r_+)|^2}{\int_0^\infty |A (-z/r_+)|^2} = \Gamma (2\Delta -1) \langle \mathcal{O}_\Delta \rangle^2 \left[ -2i\sqrt{\omega^2 - \langle V\rangle} \right]^{2(1-\Delta )} \ee
which determines $\langle V\rangle$ implicitly as a function of $\omega$.
We obtain the low-temperature high-frequency ($\omega \gtrsim \langle\mathcal{O}_\Delta\rangle^{1/\Delta}$) conductivity
\be\label{eq61} \sigma(\omega) = \sqrt{ 1 - \frac{\Gamma(2\Delta -1)\langle \mathcal{O}_\Delta\rangle^2 (-2i)^{2(1-\Delta)}}{\omega^{2\Delta}}} \ee
whereas for low frequencies, we have
\be\label{eq61a} \sigma(\omega) = \sqrt{ 1 - \frac{[2^{2(1-\Delta)} \Gamma (2\Delta -1) \langle \mathcal{O}_\Delta\rangle^2]^{1/\Delta}}{\omega^2}} \ee
In particular for $\Delta =1$ eq.~(\ref{eq60}) can be solved for all frequencies and the expression (\ref{eq61}) for the conductivity, which coincides with (\ref{eq61a}), is valid in the entire spectrum,
\be\label{eq63} \sigma(\omega) = \sqrt{ 1 - \frac{\langle \mathcal{O}_1\rangle^2}{\omega^2}} \ee
This expression is in excellent agreement with numerical results even down to low frequencies ($\omega \ll \langle\mathcal{O}_1\rangle$).

For $\Delta > 3/2$, the potential can be approximated by
\be V \approx (3-\Delta)^{2} b^2 r_+^2 (1-z^3) (b z)^{2(2-\Delta )}
\tanh^2 \left( \frac{bz}{\alpha} \right)^{2\Delta-3}
\ee
where $b$ is given in (\ref{eq37}). This expression can be used, as before, to find an estimate for $\langle V\rangle$.
In particular, for $\Delta =2$,
we obtain
\be\label{eq65} \hat{V} = 1 +2\alpha \sqrt{\hat{V} - \hat{\omega}^2} + 2\alpha^2 (\hat V - \hat\omega^2) \left[ \psi\left( \frac{\alpha}{2} \sqrt{\hat{V} - \hat{\omega}^2} \right) - \psi \left( \frac{1}{2} + \frac{\alpha}{2} \sqrt{\hat{V} - \hat{\omega}^2} \right) \right] \ee
where
\be \hat{V} = \frac{\langle V\rangle}{\langle\mathcal{O}_2\rangle\alpha} \ \ , \ \ \ \ \ \hat{\omega}^2 = \frac{\omega^2}{\langle\mathcal{O}_2\rangle\alpha} \ee
At high frequencies, this implies
\be \sigma(\omega) \approx \sqrt{ 1 + \frac{\langle \mathcal{O}_2\rangle^2}{2\omega^4}} \ee
showing that $\sigma > 1$ for $\omega \gtrsim \sqrt{\langle\mathcal{O}_2\rangle}$, whereas as $\omega\to 0$, $\hat{V} \approx 0.65$, and the low-frequency conductivity is
\be\label{eq68} \sigma(\omega) \approx 0.7i \frac{\sqrt{\langle\mathcal{O}_2\rangle}}{\omega}\ee
We shall improve on this estimate later by using a more accurate analytic technique which is better suited for low frequencies.
We shall also obtain an exponentially small real part of the conductivity which survives in the limit $\omega\to 0$.

\begin{figure}
\includegraphics{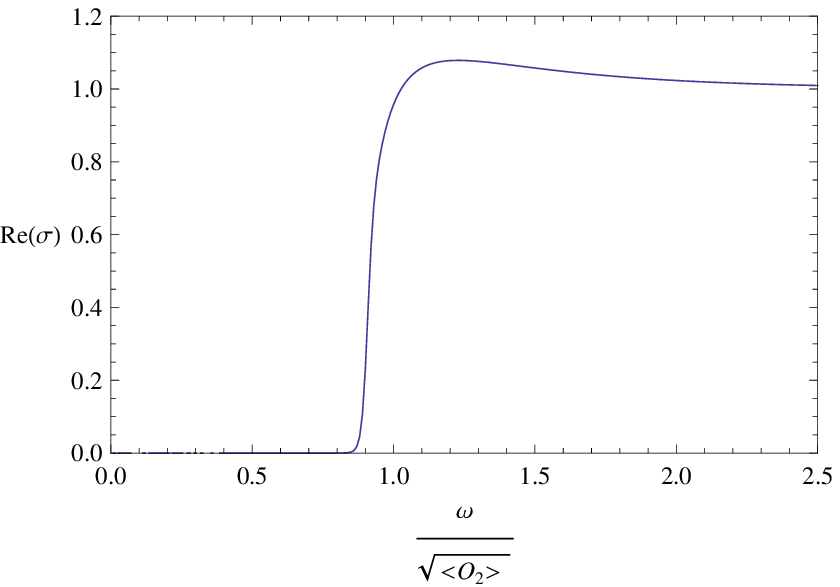}
\includegraphics{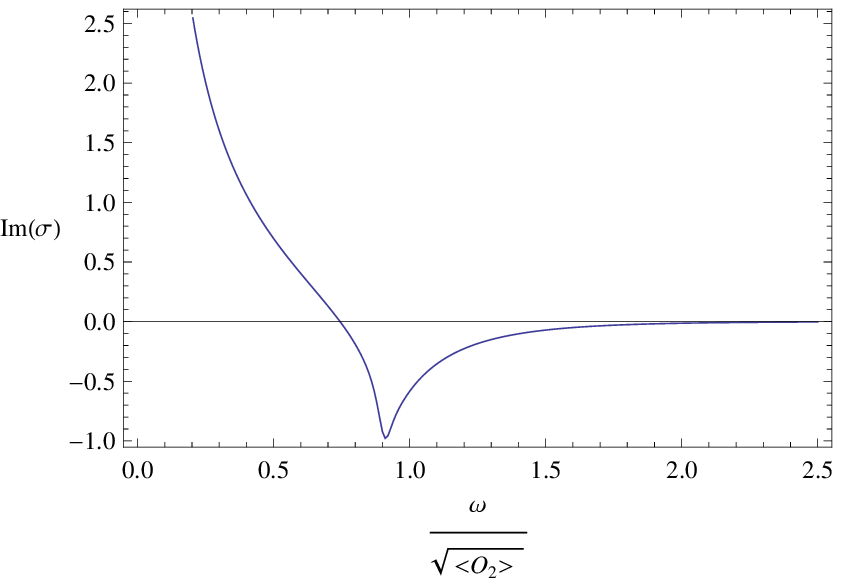}
\caption{The real and imaginary parts of the conductivity at low temperatures for $\Delta = 2$.}
\label{fig4}
\end{figure}

At intermediate frequencies, we may expand around $\hat{V} = \hat{\omega}^2 = 1$. We obtain
\be\label{eq69} \sigma(\omega) \approx \sqrt{ 1 - \frac{\alpha\langle\mathcal{O}_2\rangle}{\omega^2}}\ee
for $\omega/\sqrt{\langle\mathcal{O}_2\rangle} \approx \sqrt{\alpha} \approx 0.9$.
We may also use perturbation theory to go beyond the leading order.
Treating $\delta V = V - \langle \mathcal{O}_2\rangle \alpha$ as a perturbation,
we obtain the wavefunction
\bea A = e^{-i\sqrt{\omega^2 -\langle \mathcal{O}_2\rangle \alpha}\ r_*}
& & \left[ 1 + \frac{\alpha^2}{2\beta} - \frac{\alpha^2\pi}{\sin\beta\pi} e^{2i\sqrt{\omega^2 -\langle \mathcal{O}_2\rangle \alpha}\ r_*}
 + \frac{\alpha^2}{2\beta} F \left( \beta, 1; 1+\beta ; -e^{-2\sqrt{\langle\mathcal{O}_2\rangle/\alpha}\ r_*} \right) \right. \nonumber\\
& & \left. - \frac{\alpha^2}{2\left( 1+ \beta\right) } e^{-2\sqrt{\langle\mathcal{O}_2\rangle/\alpha}\ r_*} F \left( 1+\beta , 1, 2+\beta ; -e^{-2\sqrt{\langle\mathcal{O}_2\rangle/\alpha}\ r_*} \right)
\right] \eea
where
\be \beta = i\alpha\sqrt{ \frac{\omega^2}{\langle \mathcal{O}_2\rangle \alpha}-1}\ee
We deduce the conductivity
\be\label{eqcond} \sigma (\omega) \approx \sqrt{ 1 - \frac{\langle\mathcal{O}_2\rangle\alpha}{\omega^2}}
- \frac{i\sqrt{\langle\mathcal{O}_2\rangle\alpha^3}}{\omega} \ee
improving on the leading order expression (\ref{eq69}).
The conductivity resulting from our analytic procedure (both real and imaginary parts) is plotted in fig.~\ref{fig4} for the entire spectrum.

At low frequencies, the above expressions do not properly
account for the boundary condition at the horizon.
To this end, define
\be A = (1-z)^{-i\omega/(3r_+)} \mathcal{G} (z) \ee
where $\mathcal{G}$ is regular at the horizon ($z=1$).
The wave equation (\ref{eq53}) reads
\be\label{eq71} - 3(1-z^3)\mathcal{G}'' + (9z^2-2(1+z+z^2)\frac{i\omega}{r_+} ) \mathcal{G}'
+ \left[ \frac{6\Psi^2}{z^2} - (1+2z) \frac{i\omega}{r_+} - \frac{(2+z)(4+z+z^2)}{3(1+z+z^2)} \frac{\omega^2}{r_+^2} \right] \mathcal{G} = 0 \ee
At the horizon we may expand $\mathcal{G}$ in a Taylor series. We deduce the boundary condition
\be\label{eq58a} \mathcal{G}'(1) +\left[ \frac{2\Psi^2(1)}{3-2i\omega/r_+} - \frac{i\omega}{3r_+} \right] \mathcal{G} (1) = 0 \ee
At low temperature, for $\Delta = 1$, we have $\Psi \approx \frac{\langle \mathcal{O}_1 \rangle}{\sqrt 2 r_+} z$.
We may rescale $z\to z/b$, where $b= \langle\mathcal{O}_1\rangle /r_+$ and then let $b\to\infty$.
We obtain the approximate solution
\be A = \left( c_+ e^{+ \langle \mathcal{O}_1 \rangle^2\ z/r_+}
+ c_- e^{- \langle \mathcal{O}_1 \rangle^2\ z/r_+} \right)
e^{-\frac{i\omega z }{ 3r_+}} \ee
which is valid for low frequencies ($\omega \ll \langle\mathcal{O}_1 \rangle$).
%
We deduce the conductivity
\be \sigma (\omega) \approx i\frac{1-\frac{c_+}{c_-}}{1+\frac{c_+}{c_-}} \ \frac{\langle \mathcal{O}_1\rangle}{\omega}  \ee
The ratio $c_+/c_-$ is found by applying the boundary condition (\ref{eq58a}).
We obtain
\be \frac{c_+}{c_-} =
\frac{a-3}{a+3} e^{-2a} + \frac{2(2a^2-3)}{a(a+3)^2} e^{-2a} \frac{i\omega}{r_+} + \mathcal{O} (\omega^2)
\ \ , \ \ \ \
a = \frac{\langle\mathcal{O}_1\rangle}{r_+} \ee
We deduce the low frequency expansion
\be \sigma (\omega) = \frac{i\langle\mathcal{O}_1\rangle}{\omega} \left[ 1 - 2 \frac{a-3}{a+3} e^{-2a} - \frac{2(2a^2-3)}{a(a+3)^2} e^{-2a} \frac{i\omega}{r_+} + \mathcal{O} (\omega^2)
 \right] \ee
in agreement with the leading order result (\ref{eq63}).
The DC conductivity is
\be \Re\sigma (0) \sim e^{-2a} = e^{-\Delta_g/T} \ \ , \ \ \ \
\Delta_g = \frac{3\langle\mathcal{O}_1\rangle}{2\pi } \approx 0.48 \langle\mathcal{O}_1\rangle \ee
For $\Delta =2$, at low temperatures eq.~(\ref{eq71}) reads
\be -3\mathcal{G}'' - \frac{2i\omega}{r_+} \mathcal{G}' + \left[ 3b^2\tanh^2 \frac{bz}{\alpha} - \frac{8\omega^2}{r_+^2} \right] \mathcal{G} = 0 \ee
whose general solution is given  in terms of Legendre functions,
\be \mathcal{G} (z) \approx \left( \frac{1-\tanh \frac{bz}{\alpha} }{1+\tanh \frac{bz}{\alpha} } \right)^{\frac{i\omega\alpha}{6br_+}} \left[ c_+ P_{\frac{1}{2} (-1+\sqrt{1+4\alpha^2})}^{+\alpha} \left( \tanh \frac{bz}{\alpha} \right)
+ c_- P_{\frac{1}{2} (-1+\sqrt{1+4\alpha^2})}^{-\alpha} \left( \tanh \frac{bz}{\alpha} \right) \right]
\ee
We deduce the conductivity
\be \sigma(\omega) \approx i \frac{\sqrt{\langle\mathcal{O}_2\rangle}}{\omega}\ \frac{0.47-0.66 \frac{c_+}{c_-}}{0.85-0.30 \frac{c_+}{c_-}} \ee
The ratio $c_+/c_-$ is found from the boundary condition at the horizon (\ref{eq58a}).
At $z\approx 1$ we have $\tanh \frac{bz}{\alpha} \approx 1$, so we may approximate
\be P_{\frac{1}{2} (-1+\sqrt{1+4\alpha^2})}^{\pm\alpha} \left( \tanh \frac{bz}{\alpha} \right) = \frac{2^{\pm \alpha/2}}{\Gamma (1\mp\alpha)} \left( 1- \tanh \frac{bz}{\alpha} \right)^{\mp \alpha/2} + \dots \ee
We obtain
\be \mathcal{G} (1) \approx \left[ \frac{c_+}{\Gamma (1-\alpha)}\ e^{+b} + \frac{c_-}{\Gamma (1+\alpha)}\ e^{-b} \right] e^{-\frac{i\omega}{3r_+}} \ \ , \ \ \ \
\mathcal{G}' (1) \approx \left[ \frac{c_+\left( b -\frac{i\omega}{3r_+} \right) }{\Gamma (1-\alpha)} \ e^{+b} - \frac{c_-\left( b + \frac{i\omega}{3r_+} \right) }{\Gamma (1+\alpha)} \ e^{-b} \right] e^{-\frac{i\omega}{3r_+}} \ee
Applying the boundary condition (\ref{eq58a}), we obtain
\be \frac{c_+}{c_-} = - e^{-2b} \frac{\Gamma (1-\alpha)}{\Gamma (1+\alpha)} \left[ 1 + \frac{4i\omega}{br_+} + \dots \right] \ee
showing that at low frequencies,
\be \Im\sigma(\omega) \approx 0.55i \frac{\sqrt{\langle\mathcal{O}_2\rangle}}{\omega} \ \ , \ \ \ \ \Re\sigma(\omega) \sim e^{-2b} = e^{-\Delta_g/T} \ \ , \ \ \ \
\Delta_g \approx \frac{3\sqrt{\alpha\langle\mathcal{O}_2\rangle}}{2\pi } \approx 0.43 \sqrt{\langle\mathcal{O}_2\rangle} \ee
to be compared with our earlier estimate (\ref{eq68}) of the imaginary part which was obtained via a different, less accurate, analytic method.
The real part is exponentially small and was not detected earlier.

The above method can also be applied to other values of the dimension $\Delta$
if one replaces the potential by its self-consistent average $\langle V\rangle$.
Then by solving the wave equation (\ref{eq53}) using perturbation theory, we obtain the real part of the conductivity in the limit $\omega\to 0$ and therefore the gap $\Delta_g$ ($\Re\sigma (0) \sim e^{-\Delta_g/T}$) for all values of the dimension $\Delta$.

\section{Conclusion}
\label{sec6}

We have discussed analytic calculations involving holographic superconductors
in the probe limit \cite{Hartnoll:2008vx}. These systems are labeled by the dimension $\Delta$ of the operator that condences below a certain critical temperature $T_c$.
We found approximate explicit solutions of the non-linear field equations in the bulk near the critical temperature as well as in the zero temperature limit.
We obtained an analytic expression for the critical temperature in terms of an eigenvalue associated with the field equation of the scalar and showed that it was in good agreement with numerical results.
At low temperatures, we showed that the condensate diverges as $\langle \mathcal{O}_\Delta \rangle \sim T^{-\Delta/3}$ for $\Delta < 3/2$ signaling the breakdown of the probe approximation.
For
$\Delta > 3/2$, we obtained an expression for the condensate at zero temperature in terms of an eigenvalue associated with the field equation of the scalar and demonstrated agreement with numerical results.
Our method becomes unreliable in the limit $\Delta\to 3$. We presented evidence
that this limit is singular signaling the onset of a phase transition \cite{Horowitz:2009ij}.
We also calculated the conductivity analytically for various values of $\Delta$ and obtained good agreement with numerical results.
In the DC limit we showed that the real part of the conductivity behaves as $e^{-\Delta_g/T}$ and found analytic estimates of the gap $\Delta_g$.

It would be interesting to extend our results beyond the probe limit by including back reaction to the bulk metric. Studying the resulting field equations will enable us to take the zero temperature limit without the obstruction of numerical instabilities. This will elucidate the nature of the ground state.
Work in this direction is in progress.

\section*{Acknowledgment}

Work supported in part by the Department of Energy under grant DE-FG05-91ER40627.

\end{document}